\documentclass[aps,prb,twocolumn,superscriptaddress,floatfix]{revtex4-1}
\usepackage[colorlinks,bookmarks=false,citecolor=blue,linkcolor=red,urlcolor=blue]{hyperref}
\usepackage{epsfig}
\usepackage{tikz}
\usepackage{graphicx,comment}
\usepackage{enumitem}
\setlist[itemize]{align=parleft,left=0pt..1em}

\usepackage{dcolumn}
\usepackage{bm}

\usepackage{amsmath,amssymb}

\usepackage{bbold}

\newcommand{\nn}{\nonumber}          
\newcommand{\bea}{\begin{eqnarray}}          
\newcommand{\eea}{\end{eqnarray}}          
\newcommand{\la}{\langle}          
\newcommand{\ra}{\rangle}

\allowdisplaybreaks

\begin{document}

\title{Correlations in randomly stacked solids}
\author{Amna Khairi Nasr}
\affiliation{Sir Winston Churchill Secondary School, St Catharines, Ontario L2T 2N1, Canada}
\affiliation{Department of Physics, Brock University, St. Catharines, Ontario L2S 3A1, Canada}
\author{R. Ganesh}
\email{r.ganesh@brocku.ca}
\affiliation{Department of Physics, Brock University, St. Catharines, Ontario L2S 3A1, Canada}

\date{\today}

\begin{abstract}
Packing of spheres is a problem with a long history dating back to Kepler's conjecture in 1611. The highest density is realized in face-centred-cubic (FCC) and hexagonal-close-packed (HCP) arrangements. These are only limiting examples of an infinite family of maximal-density structures called Barlow stackings. They are constructed by stacking triangular layers, with each layer shifted with respect to the one below. 
At the other extreme, Torquato-Stillinger stackings are believed to yield the lowest possible density while preserving mechanical stability. They form an infinite family of structures composed of stacked honeycomb layers. 
In this article, we characterize layer-correlations in both families when the stacking is random. To do so, we take advantage of the H\"agg code -- a mapping between a Barlow stacking and a one-dimensional Ising magnet. The layer-correlation is related to a moment-generating function of the Ising model. 
We first determine the layer-correlation for random Barlow stacking, finding exponential decay. We next introduce a bias favouring one of two stacking-chiralities -- equivalent to a magnetic field in the Ising model. Although this bias favours FCC ordering, there is no long-ranged order as correlations still decay exponentially. Finally, we consider Torquato-Stillinger stackings, which map to a combination of an Ising magnet and a three-state Potts model. With random stacking, the correlations decay exponentially with a form that is similar to the Barlow problem. We discuss relevance to ordering in clusters of stacked solids and for layer-deposition-based synthesis methods. 
\end{abstract}

\maketitle

\section{Introduction}
Stacking is ubiquitous in solid-state materials. Of elemental solids, more than half form `close-packed' structures\cite{Steurer2001,Achim_page}. These include hexagonal close-packed (HCP), face-centred cubic (FCC), double-HCP and 9R structures. These structures are all built from stacked triangular layers. 
Stacking is also seen in non-close-packed elemental solids such as graphite, where each layer is a honeycomb lattice. Among multi-elemental solids, transition metal dichalcogenides are a stacked family. In all such solids, the underlying stacking-principles hold the key to understanding various physical properties. For example, stacking order affects x-ray diffraction\cite{Randombook_1994}, phonon spectra\cite{Cancado2008}, electronic band structure\cite{Elatresh2017} and even mechanical properties\cite{Su_2021}. In this article, we seek to understand correlations that emerge from random stacking protocols. We show that, despite randomness, the geometric constraints of stacking lead to short-ranged correlations. This acquires relevance for novel synthesis methods that proceed in a layer-by-layer fashion\cite{Gao2012,Yang2019,Nguyen2020}.

Ordering in close-packed structures has been studied extensively in theoretical models (e.g., see Ref.~\onlinecite{Jackson2002}) as well as in materials (e.g., in Ref.~\onlinecite{Loach_2017}). A key area of interest is the emergence of long-range order with periodicity in the stacking direction. This involves competition between energy and entropy contributions. When energy dominates over entropy, various ordered states emerge -- as exemplified by the celebrated ANNNI model\cite{Selke_1988,Yeomans1988}. When energies are comparable, entropy can select a particular ordered state, e.g., see Ref.~\onlinecite{Woodcock1997}. In this article, we discuss the nature of correlations in a purely entropic setting. We work within a rigid-layer picture where geometry constrains the relative position of neighbouring layers. We address a  particularly simple question: what is the probability that two distant layers, separated by $N$ intervening layers, are aligned? We answer this question for three different stacking schemes below.

\section{Barlow stackings as a 1D Ising model}

Atoms within a solid can be modelled as spheres that are packed in three dimensional space. In 1611, Kepler conjectured that the highest possible density occurs in a face-centred cubic arrangement\cite{Kepler_1611}. This was rigorously demonstrated as recently as in 2017\cite{Hales_2017}. It has long been known that the FCC lattice is just one among an infinite number of arrangements, all with the same density\cite{TS2010}. This family of close-packed structures, also called Barlow stackings\cite{Barlow_1883}, is constructed with triangular layers as building blocks. When one layer is stacked upon another, it must be laterally shifted along one of two directions. This leads to three possible positions for each layer, denoted as A, B and C. In order to ensure maximal density, we must not repeat letters in succession. We then have a two-fold choice at each layer. The number of configurations grows exponentially with the number of layers.

Consider a Barlow stacking with $(N+1)$ layers. It can be expressed as a string of letters $L_1 L_2 \ldots L_{N+1}$, where each $L_i$ takes one of three values: $A$, $B$ or $C$. The only constraint is that any two adjacent letters cannot be the same, e.g., $ABCCA\ldots$ is forbidden while $ABCABC\ldots$ is allowed. Any such structure can be coded as a string of $N$ Ising variables via a dual construction, known as the H\"agg code\cite{Hagg1943}. We represent it as $\sigma_1 \sigma_2\ldots \sigma_N$ where each $\sigma_i$ takes one of two values, $\pm 1$. The variable $\sigma_i$ represents the shift when moving from layer $i$ to layer $(i+1)$. A forward shift, i.e., $A \rightarrow B$, $B\rightarrow C$ or $C\rightarrow A$, is encoded as $\sigma = +1$. A backward shift, i.e., $A \rightarrow C$, $B\rightarrow A$ or $C\rightarrow B$, is encoded as $\sigma = -1$.

We define the layer correlation function $P_N^{\mathrm{B.}} $ as the probability that layer $i$ and layer $i+N$ are aligned, i.e., $P_N^{\mathrm{B.}} = P(L_i = L_{i+N})$. Here, $B.$ stands for Barlow. For example, if the first layer is $A$, $P_{N}^{\mathrm{B.}}$ is the probability that the $(N+1)^\mathrm{th}$ layer is also $A$. 
This quantity encodes memory of the initial layer as stacking progresses. In the following sections, we will evaluate $P_N^{\mathrm{B.}}$ for two stacking schemes. It is convenient to rephrase this quantity in the language of Ising variables. From layer $i$ to layer $(i+N)$, we encounter $N$ Ising variables. There are $2^N$ possible configurations of these Ising variables, representing all possible configurations of intervening layers. In a particular stacking configuration, suppose $P$ Ising variables have value $+1$ and $Q$ have $-1$. The sum $P+Q = N$ is fixed. Consider $A$, $B$ and $C$ to be arranged in a circle (to have periodic boundaries). As each additional layer is deposited, we move forward or backward (clockwise or counter-clockwise) along the circle. The net number of steps in the forward direction is $\sum_{i=1}^N \sigma_i= P-Q$. The $(i+N)^\mathrm{th}$ layer will be the same as the $i^\mathrm{th}$  if the net number of steps is a multiple of three. That is, $L_{i+N} = L_i$ if $(P - Q)~ \mathrm{mod} ~3 = 0$. 

To evaluate the corresponding likelihood, we divide the $2^N$ Ising configurations into three classes based on $(P - Q)~ \mathrm{mod} ~3$. We denote the likelihood of $(P - Q)~ \mathrm{mod} ~3 = 0$ as $\Pi_0(N)$, given by
\bea
\Pi_0(N) = \sum_{\sigma_1,\ldots,\sigma_N} p(\sigma_1,\ldots,\sigma_N) \delta\Big( \big\{\sum_{i=1}^N \sigma_i\big\} ~\mathrm{mod}~3,0 \Big).~~~~
\eea
Here, the sum over $(\sigma_1,\ldots,\sigma_N)$ amounts to summing over all Ising configurations. The likelihood of any particular configuration is denoted by $p(\sigma_1,\ldots,\sigma_N)$. Finally, the delta function selects configurations where the Ising-sum is a multiple of three (i.e., with $\big\{\sum_{i=1}^N \sigma_i\big\} ~\mathrm{mod}~3 =0$). On the same lines, we define
\bea
\Pi_1(N) = \sum_{\sigma_1,\ldots,\sigma_N} p(\sigma_1,\ldots,\sigma_N) \delta\Big( \big\{\sum_{i}\sigma_i\big\}~\mathrm{mod}~3,1 \Big),~~~~\\
\Pi_2(N) = \sum_{\sigma_1,\ldots,\sigma_N} p(\sigma_1,\ldots,\sigma_N) \delta\Big(\big\{\sum_{i} \sigma_i\big\}~\mathrm{mod}~3,2\Big).~~~~
\eea
As there are only three possibilities, the total probability is given by
\bea
\Pi_0(N) + \Pi_1(N) +\Pi_2(N) = 1.
\label{Eq.totprob}
\eea
We now consider $e^{i\Omega \sum_{i=1}^N \sigma_i}$, where $\Omega = 2\pi/3$. In configurations with $(P - Q)~ \mathrm{mod} ~3 = 0$, this quantity is unity. In the other two classes, $e^{i\Omega \sum_{i} \sigma_i} = \frac{-1}{2}\pm i\frac{\sqrt{3}}{2}$. We have
\bea
\la e^{i \Omega \sum_{i} \sigma_i} \ra = \Pi_0 (N)+ \Pi_1(N) e^{i2\pi/3} + \Pi_2 (N) e^{i4\pi/3}.~~~~
\label{eq.2}
\eea

The layer-correlation $P_N^{\mathrm{B.}}$ is simply the probability that $(P - Q)~ \mathrm{mod} ~3 =0$. In other words, $P_N^{\mathrm{B.}} = \Pi_0 (N)$. From Eqs.~\ref{Eq.totprob} and \ref{eq.2}, we express it as
\bea
P_N^{\mathrm{B.}}= \Pi_0 (N) = \frac{1}{3} \Big(
1 + 2~   \mathrm{Re}\Big\{\Big\la e^{i \Omega \sum_{i} \sigma_i} \Big\ra\Big\} 
\Big),
\label{eq.Pmomgen}
\eea 
where $\mathrm{Re}\{\cdot \}$ represents the real part. This relation ties the layer-correlation function to a certain moment-generating function of the Ising model, $\Big\la e^{i \Omega \sum_{i} \sigma_i} \Big\ra$. We emphasize that Eq.~\ref{eq.Pmomgen} holds for any Barlow stacking, i.e., for any close-packed structure. In the following sections, we will explicitly evaluate this layer-correlation function for two stacking schemes. 

For later use, we also use Eqs.~\ref{Eq.totprob} and \ref{eq.2} to write
\bea
 \Pi_1 (N) = \frac{1}{3} \Big(
1 + 2~   \mathrm{Re}\Big\{\Big\la e^{i \Omega \big\{2+\sum_{i} \sigma_i\big\}} \Big\ra\Big\} 
\Big), \label{eq.Pmomg1} \\
 \Pi_2 (N) = \frac{1}{3} \Big(
1 + 2~   \mathrm{Re}\Big\{\Big\la e^{i \Omega \big\{1+\sum_{i} \sigma_i\big\}} \Big\ra\Big\} 
\Big).
\label{eq.Pmomg}
\eea

\subsection{Random Barlow stacking}
\label{sec.randBar}

As each layer is deposited, it is assumed to randomly select one of two allowed positions. In the Ising language, this corresponds to generating a set $N$ Ising variables at random, e.g., by flipping a coin $N$ times. At each flip, the two outcomes are equally likely. This leads to a truly random sampling with each Ising configuration having the same likelihood, 
\bea
p(\sigma_1,\ldots,\sigma_N) = \frac{1}{2^N}.
\eea
To evaluate the layer-correlation function of Eq.~\ref{eq.Pmomgen}, we consider
\bea
\nn \frac{ (e^{i\Omega} + e^{-i\Omega})^N}{2^N} =\frac{1}{2^N} \sum_{\sigma_1,\ldots,\sigma_N} e^{i\Omega \{\sum_{i=1}^N \sigma_i\}} \\
= \la e^{i\Omega \sum_i \sigma_i} \ra.
\label{eq.4}
\eea
Here, we have used the standard binomial expansion. In the summation over $\sigma_1, \ldots, \sigma_N$, each variable runs over the two values $\pm 1$. We have interpreted the expression on the right as an expectation value over configurations of $N$ Ising variables.  
At the same time, we have $e^{i\Omega} + e^{-i\Omega} = -1$, from the explicit expressions for $e^{\pm i\Omega}$. We arrive at $\la e^{i\Omega \sum_i \sigma_i} \ra = (-1)^N/2^N$. From Eq.~\ref{eq.Pmomgen}, we now obtain
\bea
P_N^{(\mathrm{r.~B.})} = \frac{1}{3} \left(
1 + \frac{(-1)^N}{2^{N-1}} 
\right),
\label{eq.Prandom}
\eea
where $\mathrm{r.~B.}$ stands for random Barlow. This layer-correlation function has two pieces: an $N$-independent contribution of $\frac{1}{3}$ and a term that decays as $\sim 2^{-N}$. Notably, the latter term is oscillatory in character as it switches sign between even and odd values of $N$. At $N=1$, $P^{(\mathrm{r.~B.})}$ vanishes as two successive layers cannot be the same. At $N \rightarrow \infty$, the oscillatory term vanishes and we are left with $P_N \rightarrow 1/3$. This encodes the fact that at large separations, each layer takes one of three values: $A$, $B$ or $C$. These three layer positions are sampled uniformly so that the probability of the final layer aligning with the first is $1/3$. Essentially, the system has no memory of the initial state. This is not the case at short distances, where $P_N$ oscillates about $1/3$. 

\subsection{Biased Barlow stacking}
\label{ssec.biasedBarlow}
We have evaluated the layer correlation function for random stacking using the Ising representation. In terms of Ising variables, each successive Ising variable was taken to be $+1$ or $-1$ with equal likelihood. We now introduce a bias that is equivalent to a magnetic field in the Ising language, with a higher likelihood for one of the Ising variables (say $+1$). This is equivalent to favouring one stacking `chirality' -- if a certain layer is $A$, the next is more likely to be $B$ rather than $C$.  

We take the probability for $+1$ to be $\alpha$ and that for $-1$ to be $(1-\alpha)$. Each configuration of Ising variables is assigned a probability $p(\sigma_1,\ldots, \sigma_N) = \alpha^{P} (1-\alpha)^{Q}$. Here, $P$ is the number of Ising variables with value $+1$ while $Q$ represents the number of $-1$'s. We may now write
\bea
\la e^{i\Omega  \sum_i \sigma_i} \ra = \sum_{\sigma_1,\ldots,\sigma_N} \alpha^{P} (1-\alpha)^{Q} e^{i \Omega \sum_i \sigma_i}.
\eea
To obtain a closed form, we rewrite $\sum_i \sigma_i = (P-Q)$. We then have
\begin{eqnarray}
\nonumber \la e^{i\Omega \sum_i \sigma_i} \ra &=& \sum_{\sigma_1,\ldots,\sigma_N}  \alpha^{P} (1-\alpha)^{Q}e^{i\Omega P - i \Omega Q }  \\
 &=& \big(\alpha e^{i\Omega} + (1-\alpha)e^{-i\Omega}\big)^N.
\eea
At the last step, we have used the binomial expansion. We may rewrite $\big(\alpha e^{i\Omega} + (1-\alpha)e^{-i\Omega}\big)$ as one complex number, $ z= \zeta e^{i\theta}$, with amplitude $\zeta$ and phase $\theta$. We have 
\bea
\zeta  &=& \sqrt{1+3\alpha^2-3\alpha},\\
\theta &=& \left\{ \begin{array}{c}
\pi -  \tan^{-1}\big(\sqrt{3}(2\alpha-1)\big),~~\alpha > \frac{1}{2} \\
\tan^{-1}\big(\sqrt{3}(1-2\alpha)\big),~~\alpha < \frac{1}{2} \end{array}\right.
\eea
In terms of these quantities, we have
\bea
\mathrm{Re}\{ \la e^{i\Omega  \sum_i \sigma_i} \ra \}=  \zeta^N \cos (N\theta).
\eea 
From Eq.~\ref{eq.Pmomgen}, the layer correlation function comes out to be
\bea
P_N^{\mathrm{b.~B.}} = \frac{1}{3} \Big(
1+ 2~ \zeta^N \cos (N\theta)
\Big),
\label{eq.Pbiased}
\eea
where $\mathrm{b.~B.}$ stands for biased Barlow. 
As with random Barlow stacking, the layer correlation carries an $N$-independent contribution of $\frac{1}{3}$ and a term that decays exponentially (note that $\zeta \leq 1$). For large $N$, the probability approaches $\frac{1}{3}$, reflecting the number of possible layer positions. There is no memory of the initial position as positions A, B and C are sampled uniformly. For small $N$, however, there is an oscillatory correction due to the cosine term. Unlike the random case, the period of the oscillation depends on $\alpha$. When $\alpha$ is strictly zero or strictly unity, we obtain an ordered FCC structure with $-ACB-$ or $-ABC-$ stacking. At these two limiting values, $\zeta$ approaches unity while $\theta$ approaches $\pm 2\pi/3$. The resulting $P_N$ has a periodicity of three, with $P_1 =0$,  $P_2 =0$, $P_3 =1$, $P_4 =0$, $P_5 =0$, $P_6 =1$, etc.

 \section{Torquato-Stillinger stackings and their dual representation}
   
Torquato-Stillinger (TS) stackings are built from honeycomb layers\cite{TS2007,TS2010,Torquato2018,Burnell2008}. They can be viewed as derivatives of Barlow stackings, with one-third of the spheres removed from each layer. They form `tunnelled' crystals, with tunnels carved through a Barlow stacking framework. As described in previous sections, Barlow stackings can be characterized as follows: a two-fold choice at each step ($\sigma = \pm 1$) leading to three possible positions for each layer ($A$, $B$ or $C$). In direct analogy, TS packings correspond to a six-fold choice at each step. In turn, this leads to nine possible positions for each layer. This can be understood as follows. 
 
\begin{figure}
\includegraphics[width=3.2in]{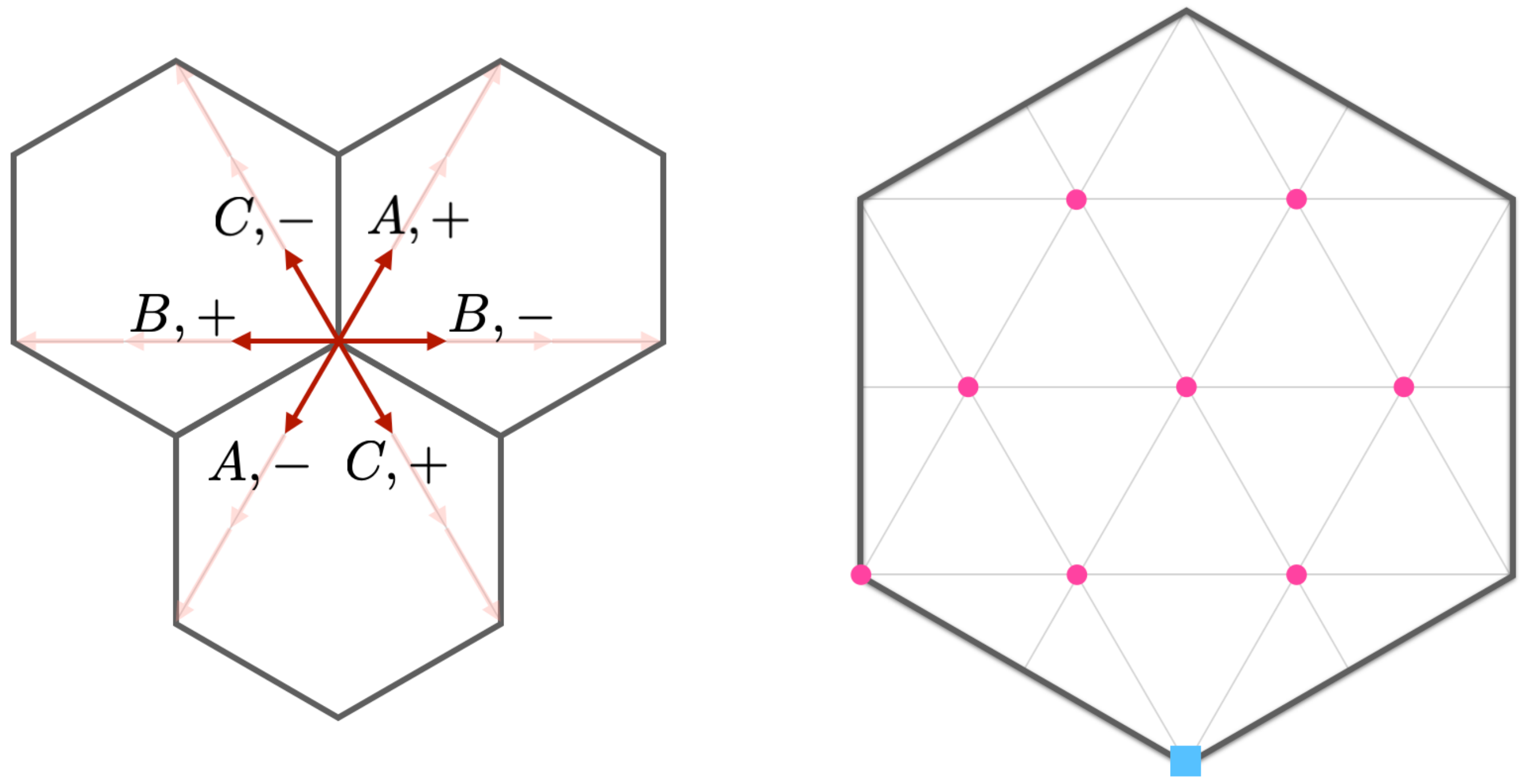}
\caption{Torquato-Stillinger stackings. Left: Each layer must be laterally displaced with respect to the previous one. The lateral displacement vector can be any one of the six possible choices shown. Each choice is one-third of a lattice translation vector of the honeycomb lattice. That is, with three displacements by the same vector, there is no net lateral displacement. Right: We show the nine possible lateral positions of layers. We mark a reference site on the base layer (shown as a blue square). In a future layer, the corresponding point can be laterally shifted to one of the eight positions shown (magenta circles). It may also be aligned with reference site (blue square).  }
\label{fig.TS}
\end{figure}

Each honeycomb layer is displaced with respect to the preceding layer by one of six possible lateral displacement vectors\cite{Burnell2008}. These vectors have the same magnitude, but are uniformly spread in direction as shown in Fig.~\ref{fig.TS}. As each layer is deposited, a six-fold choice is made. For our purposes, it is convenient to picture this as one threefold-choice of direction ($d = A,~B,~C$) and one two-fold choice of step ($\sigma = \pm 1$). The three-fold choice is depicted as A, B and C directions in the figure, while the step is denoted with $\pm$. Once a direction is chosen, we may move along or opposite -- corresponding to the two-fold choice encoded in $\sigma$. 
 
Before broaching layer correlations, we note that the honeycomb lattice has two primitive lattice vectors. A honeycomb layer is unchanged by a displacement by any integer combination of these primitive lattice vectors. As seen Fig.~\ref{fig.TS}, three consecutive displacements along the same direction (any one of the six allowed directions) aligns the honeycomb lattice with itself. In fact, we may displace a layer along each direction ($A$, $B$ or $C$) by a different multiple of three. This will result in the same final position for the layer.

Labelling the layers as $M_1 M_2 \cdots$, we may compare the relative positions of layer $M_1$ and $M_{N+1}$. To reach $M_{N+1}$ from $M_{1}$, we must add $N$ lateral displacement vectors (apart from $N$ copies of the vertical stacking vector). This sum must be evaluated modulo the primitive lattice vectors. By examining all vector combinations, we see that the sum can only take nine values as shown in Fig.~\ref{fig.TS} (right).

We now define the layer correlation function, $P_N^{\mathrm{TS}}$, as the probability that $M_1$ and $M_{N+1}$ are laterally aligned. We may separate the $N$ intervening lateral displacements into three classes associated with the three directions $A$, $B$ and $C$. For $M_1$ and $M_{N+1}$ to be aligned, one of the following three conditions must be satisfied:

\begin{itemize}
\item[i.] Along each of the three directions, net displacement (in units of step length) must be a multiple of 3.  
\item[ii.] Along each direction, net displacement must be of the form $(3n+1)$, where $n$ is any integer. 
\item[iii.] Along each direction, net displacement must be of the form $(3n+2)$, where $n$ is any integer. 
\end{itemize}

These three cases can be understood from Fig.~\ref{fig.TS} (right). By explicitly adding various net displacements, we see that these three cases result in zero-net-lateral-displacement. In all other cases, the final layer is shifted with respect to the first.  

Below, we calculate the layer correlation function by adding the probabilities of these three cases.

 \subsection{Random Torquato-Stillinger stacking}
With each layer in a TS stacking, a six-fold choice is made. Here, we assume that the six possibilities are equally likely. 
That is, when a layer is deposited, each of the three directions is equally likely. The two values of the step variable are also equally likely. 

Suppose $N_A$ steps were taken in the $A$ direction (including forward and backward steps), $N_B$ in the $B$ direction and $N_C$ in the $C$ direction. We must have
\bea
N_A + N_B + N_C = N.
\eea 
Restricting our attention to displacements in the $A$ direction, we have an effective random-Barlow-stacking problem with $N_A$ layers. 
The net displacement in the $A$ direction may be zero, one or two (as displacement is calculated modulo three). The probability for zero displacement is $P_{N_A}^{(\mathrm{r.~B.})}$, where $P^{(\mathrm{r.~B.})}$ is defined in Eq.~\ref{eq.Prandom} above. In the same manner, the probabilities for net zero displacements along $B$ and $C$ are $P_{N_B}^{(\mathrm{r.~B.})}$ and $P_{N_C}^{(\mathrm{r.~B.})}$ respectively.
 
The joint probability for all three directions to have net-zero-displacement is 
\bea
\nonumber P_{0,0,0} &=& \frac{1}{3^N} \sum_{N_A + N_B + N_C = N} \frac{N !}{N_A ! N_B ! N_C !}\times \\
&~& P_{N_A}^{(\mathrm{r.~B.})} \times P_{N_B}^{(\mathrm{r.~B.})}\times P_{N_C}^{(\mathrm{r.~B.})}.~~~~~~
\eea
Here, the sum over $N_A$, $N_B$ and $N_C$ represents all three-partitions of $N$. That is, it runs over all (non-negative integer) values of $N_A$ and $N_B$ and $N_C$ with the constraint that they must add to $N$. The term $\frac{N !}{N_A ! N_B ! N_C !}$ accounts for all possible reorderings of the direction variables. In this expression, the $N$ step variables ($\sigma_i = \pm 1$) do not appear explicitly. They are implicitly accounted for within the probabilities $P_{N_{A/B/C}}^{(\mathrm{r.~B.})}$.

This expression can be rewritten using the explicit form of $P^{(\mathrm{r.~B.})}$ from Eq.~\ref{eq.Prandom}. As shown in the appendix, each term in the resulting sum can be reexpressed as a trinomial expansion and evaluated. We find
\bea
P_{0,0,0} = \frac{1}{27} \Big[
1 + \frac{3}{2^{N-1}}+\frac{(-1)^N}{2^{N-3}}
\Big].
\eea
To evaluate the layer correlation, we also require the probabilities for cases (ii) and (iii) listed above. Case (ii) requires the net displacement in each direction to be 1 mod 3. Case (iii) requires net displacements of 2 mod 3. To find these, we revert to the problem of random Barlow stacking. With $N$ Ising variables, the probability of net displacement being 1 mod 3 is given by $\Pi_1(N)$ of Eq.~\ref{eq.Pmomg1}, that for 2 mod 3 is given by $\Pi_2(N)$ of Eq.~\ref{eq.Pmomg}. Assuming random Barlow stacking, we may follow the arguments in Sec.~\ref{sec.randBar} above to find
\bea
 \Pi_1 ^{(\mathrm{r.~B.})}(N) =\Pi_2^{(\mathrm{r.~B.})}(N) 
= \frac{1}{3} \left(
1 - \frac{(-1)^N}{2^{N}} \right). 
\eea 
For the TS stacking, we now find the probabilities for cases (ii) and (iii),   
\bea
\nn P_{1,1,1} &=& \frac{1}{3^N} \sum_{N_A + N_B + N_C = N} \frac{N !}{N_A ! N_B ! N_C !}\times \\
&~& \Pi_1^{(\mathrm{r.~B.})}(N_A) \times 
\Pi_1^{(\mathrm{r.~B.})}(N_B) \times\Pi_1^{(\mathrm{r.~B.})}(N_C),~~~~\\
\nn P_{2,2,2} &=& \frac{1}{3^N} \sum_{N_A + N_B + N_C = N} \frac{N !}{N_A ! N_B ! N_C !}\times \\
&~& \Pi_2^{(\mathrm{r.~B.})}(N_A) \times 
\Pi_2^{(\mathrm{r.~B.})}(N_B) \times\Pi_2^{(\mathrm{r.~B.})}(N_C).~~~~
\eea
Here, $ P_{1,1,1}$ represents the probability for each direction to have a net displacement of 1 mod 3. Similarly, $ P_{2,2,2}$ is the probability for each direction to have a net displacement of 2 mod 3. 
Relegating details to the appendix, we find
\bea
 P_{1,1,1} =  P_{2,2,2} =   \frac{1}{27} \Big[
 1 - \frac{3}{2^N} -\frac{(-1)^N}{2^N}
 \Big].
 \eea
The net probability that layer 1 and layer $N+1$ are aligned is given by
\bea
P_N^{(\mathrm{r.~TS})} &=& P_{0,0,0} +  P_{1,1,1} +  P_{2,2,2} =   \frac{1}{9} \Big[
 1 +  \frac{(-1)^N}{2^{N-1}}
 \Big],~~~~
 \label{eq.PTS}
\eea
where $\mathrm{r.~TS}$ denotes random TS stacking. We have arrived at the layer correlation function. This expression is very similar to the result for random Barlow stacking in Eq.~\ref{eq.Prandom}. As with the Barlow case, we obtain a constant contribution of $1/9$ and an exponentially decaying oscillatory term. With $N=1$, $P^{(\mathrm{r.~T.S})}$ vanishes as two successive layers cannot be the same. For large $N$, the layer correlation is $1/9$, reflecting the fact that there are $9$ possible lateral positions. This indicates that, over large distances, the nine lateral positions are sampled uniformly. At small distances, the probability has an oscillatory component that retains memory of the initial layer.

The above arguments can be easily modified to include biases. We may consider a bias favouring displacements along one of the three directions and/or one favouring a particular step value. Irrespective of any bias, the underlying structure will lead to exponential decay.

\section{Discussion}
The first key result of this work concerns the layer-correlation function of any Barlow stacking. It is given by a moment-generating function of a 1D Ising model. This result is similar to the work of Pandey and Krishna\cite{Pandey1977} in the context of X-ray diffraction peaks in disordered 2H structures. They evaluate the same moment-generating function for certain fault-probabilities, with the goal of finding X-ray intensities. Here, we show that this function encodes the layer-correlation function. 

The second key result concerns the explicit form of layer-correlations in random-stacking models, described in Eqs.~\ref{eq.Prandom}, \ref{eq.Pbiased} and \ref{eq.PTS} above. The three stacking models discussed here show exponential decay with similar functional forms. The exponential decay arises from entropic reasons. If the initial and final (first and $(N+1)^\mathrm{th}$) layers are fixed, the intervening layers can be in $\sim c^N$ configurations. Here, $c$ is the number of allowed positions per layer ($c=2,~6$ for random Barlow and TS stackings respectively). For large $N$, this number is approximately the same for any choice of initial and final layers. The probability of the layers aligning is the ratio of the number of configurations with aligned layers to the total number for all cases. This ratio decays exponentially as $N$ increases.

Our discussion of Barlow stackings may be relevant to materials with weak (non-covalent) inter-layer interactions. For example, noble gas solids\cite{Pollack1966} have weak inter-atomic interactions of the van der Waals type. As a result, they may not have a strong preference for a certain local coordination geometry. They are known to form close-packed structures, HCP and FCC in particular\cite{Dewaele2021}. Thermal fluctuations or disorder may lead to a large degree of stacking randomness, where our results may apply. Our discussion may also be relevant to graphite which is a non-close-packed material with honeycomb layers. However, it follows the same stacking rule as Barlow packings. Graphite is known to occur in two forms: Bernal stacking (AB) and rhombohedral stacking (ABC). As the layers are held together by weak van der Waals' bonding, it is conceivable that a high degree of stacking-randomness may occur. Recent studies have explored layer-by-layer synthesis of graphite variants\cite{Latychevskaia2018,Yang2019,Nguyen2020,Bouhafs2021}. Our results could be of relevance here.

We have discussed a model of biased Barlow stacking in Sec.~\ref{ssec.biasedBarlow} above, where one stacking chirality is favoured over another. This model can be realized in multi-element compounds where the stacking unit consists of multiple triangular layers, e.g., in recent experiments on transition metal dichalcogenides\cite{Zhao2022}. Chemical vapour deposition, under suitable conditions, may favour 3R stacking\cite{Wang2017,Ryu2019,Deng2020}, a structure that is equivalent to FCC structure. As each layer is deposited, there is a strong preference for one chirality over another. Our results can be used to quantify stacking bias solely from measurements on the initial and final layers.  

We have also discussed Torquato-Stillinger packings, a class of structures with no known material realization so far. Future studies of stacked honeycomb materials may realize this model, and thereby realize the lowest-density stable solid. Our results help to understand the large configuration entropy within this family.

\acknowledgments
We thank Zak Mason for help with 3D printing model layers. This work was supported by the Brock FMS Science Mentorship programme and a Discovery Grant 2022-05240 from the Natural Sciences and Engineering Research Council of Canada.

\appendix

\section{Evaluating layer correlations in TS stacking}
Using the explicit form of $P^{(\mathrm{Barlow})}$ from Eq.~\ref{eq.Prandom}, we obtain
\bea
\nonumber P_{0,0,0} &=& \frac{1}{27}\frac{1}{3^N} \sum_{N_A + N_B + N_C = N} \frac{N !}{N_A ! N_B ! N_C !}\times \\
\nn   &~&\Big[ 1+
\frac{(-1)^{N_A}}{2^{N_A-1}}+ \frac{(-1)^{N_B}}{2^{N_B-1}}+ \frac{(-1)^{N_C}}{2^{N_C-1}}\\
\nn &~&+ \frac{(-1)^{N_A+N_B}}{2^{N_A+N_B-2}} + \frac{(-1)^{N_B+N_C}}{2^{N_B+N_C-2}}
+ \frac{(-1)^{N_C+N_A}}{2^{N_C+N_A-2}}\\
&~&+ \frac{(-1)^{N_A+N_B+N_C}}{2^{N_A+N_B+N_C-3}}
\Big].
\eea
Each term in this expression is a straightforward example of a trinomial expansion. We have
\bea
 \sum_{N_A + N_B + N_C = N} \frac{N !}{N_A ! N_B ! N_C !} &=& 3^N,\\
 \sum_{N_A + N_B + N_C = N} \frac{N !}{N_A ! N_B ! N_C !} \frac{(-1)^{N_A}}{2^{N_A-1}} &=& \frac{3^N}{2^{N-1}},~~~~~\\
\sum_{N_A + N_B + N_C = N} \frac{N !}{N_A ! N_B ! N_C !}  \frac{(-1)^{N_A+N_B}}{2^{N_A+N_B-2}}  & =&0,\\
\sum_{N_A + N_B + N_C = N} \frac{N !}{N_A ! N_B ! N_C !}  \frac{(-1)^{N_A+N_B+N_C}}{2^{N_A+N_B+N_C-3}}
&=&\frac{(-3)^N}{2^{N-3}}.~~~~~~
\eea

On the same lines, the expression for $P_{1,1,1}$ comes out to be
\bea
\nonumber P_{1,1,1} &=& \frac{1}{27} \frac{1}{3^N}\sum_{N_A + N_B + N_C = N} \frac{N !}{N_A ! N_B ! N_C !} \times \\
\nn &~& \Big[ 1- 
\frac{(-1)^{N_A}}{2^{N_A}}- \frac{(-1)^{N_B}}{2^{N_B}}- \frac{(-1)^{N_C}}{2^{N_C}}\\
\nn &~&+ \frac{(-1)^{N_A+N_B}}{2^{N_A+N_B}} + \frac{(-1)^{N_B+N_C}}{2^{N_B+N_C}}
+ \frac{(-1)^{N_C+N_A}}{2^{N_C+N_A}}\\
&~&- \frac{(-1)^{N_A+N_B+N_C}}{2^{N_A+N_B+N_C}}
\Big].
\eea
The expression for $ P_{2,2,2}$ is identical to that of $P_{1,1,1}$.

\bibliographystyle{apsrev4-1} 
\bibliography{stacking}
\end{document}